\documentstyle[preprint,revtex]{aps}
\begin{document}
\draft
\preprint{26 Feb 1993}

\begin{title}

Spontaneous Spin-Polarization of Ballistic Electrons \\
in Single Mode
Quantum Wires due to Spin-Splitting

\end{title}

\author{Gerhard Fasol and Hiroyuki Sakaki}

\begin{instit} Quantum Wave Project (ERATO), JRDC,\\ Keyaki House 302, 4-3-24
Komaba, Meguro--ku, Tokyo 153, Japan\\ and\\ Research Center for Advanced
Science and Technology (RCAST)\\ University of Tokyo, 4-6-1 Komaba,
Meguro-ku, Tokyo 153, Japan
\end{instit}

\begin{abstract}

We show that a quantum wire device with spin splitting  can work as an active
 spin polarizer.   Hot electrons  in one `spin' subband (e.g. `spin-up') may
pass such a device with weak electron  pair scattering, while electrons in
the opposite subband (`spin-down') may have high conversion probability into
the `spin-up' subband, resulting in  spin polarization of a hot electron
beam. Under different circumstances a hot electron beam passing through a
single mode quantum wire may induce a steady state magnetization of the
background electron gas in a section of the wire weakly coupled to the
environment.

\end{abstract}


\narrowtext

In the present letter we predict two related new effects, achievable in
devices based on   single mode quantum wires  with spin splitting of the
electron bands: (1) a quantum wire structure, acting  as an active spin
polarizer for hot electrons, and (2) a device structure where a beam of hot
electrons  induces a steady state magnetization of the background electron
gas in a section of the quantum wire.  These effects allow the construction
of several new devices and experiments. One possibility is the exploration of
the spin resolved band structure and magnetic properties of quantum dots and
other microstructures.

Transport in electron wave guides and in quantum wires is presently an active
area of research (for a review see Ref.~\cite{beenakker}).  It is expected,
that the transition from two-dimensional (2D) to single mode one-dimensional
(1D) systems will show even more dramatic effects than the step from three
dimensions (3D) to 2D. Two interesting effects are predicted here. We are
here mainly concerned with  transport in narrow single mode high mobility
quantum wires (i.e. with a  width around $10 nm$) and carriers with an excess
energy $\Delta$ (typically a few meV) above the Fermi energy, although we
expect that the present results can be more generally applied.

Fig.~\ref{fig0} introduces  electron pair scattering processes in  `single
mode' quantum wires. This type of process breaks the electron phase. Although
not flipping electron spin directly, it acts effectively like  a spin flip
process. The scattering partners must be in different `spin' subbands --- in
a strictly 1D quantum wire, first order electron pair scattering is forbidden
for partners in the same spin subband. (In 2D, for comparison,  pair
scattering for electrons in the same spin subband is not forbidden, but
expected to be about 50\% weaker than for pairs with partners in opposite
spin subbands \cite{fasolssc}).

The bands in bulk semiconductors lacking inversion symmetry are in general
spin-split in zero magnetic field, with splitting terms proportional to
$|{\bf k}|$ and $|{\bf k}|^{3}$ \cite{cardona}. In group III-V quantum wells
or hetero-structures there are additional spin splitting terms due to
microscopic electric fields and confinement \cite{roessler} \cite{bastard}.
Spin splitting in 2D has recently been experimentally demonstrated by Raman
scattering \cite{jusserand} and in transport \cite{das} \cite{dresselhaus}.
Fascinating new transport effects due to spin splitting have been predicted
recently \cite{goldoni}. Splitting of the spin subbands is also expected for
quantum wires. In quantum wires, the wave functions of the two spin split
subbands will have mixed character. We label `spin-up' and `spin-down' bands
here according to the major contribution.

Fig.~\ref{fig1}(a) shows a typical electron-electron pair scattering process
in a quantum wire with spin-splitting. An electron in the `spin-up' subband
at $\left( {\bf p}_{1}, \uparrow \right)$ scatters with a spin-down electron
at $\left( {\bf k}_{1}, \downarrow \right)$, resulting in a hole at $\left(
{\bf k}_{1}, \downarrow \right)$, and  electrons  at $\left( {\bf k}_{1} -
{\bf q}_{1}, \downarrow \right)$ and  at  $\left( {\bf p}_{1} + {\bf q}_{1},
\uparrow \right)$. This process has a much lower probability than a similar
process for an electron in the `spin-down' subband, because the final state
$\left( {\bf p}_{1}  + {\bf q}_{1}, \uparrow \right)$ has a high probability
of being occupied, while the state ${\bf k}_{1}$ has a high probability of
being empty. Therefore, an electron injected into the `spin-up' subband is
likely to pass the quantum wire without scattering. An electron $\left( {\bf
k}_{2} - {\bf q}_{2}, \downarrow \right)$ injected into the `spin-down'
subband, on the other hand, has an increased scattering probability as shown
in Fig.~\ref{fig1}(b). There is a high probability that an injected electron
in the `spin-down' subband is converted into a hot `spin-up' subband electron
at similar energy. Therefore the quantum wire can act as an active electron
polarizer. This effect relies on the  ${\bf k}$ dependence of the spin
splitting, mentioned above. The polarizer effect  survives mixing of the spin
subband wave functions due to confinement.

For our calculation we express the scattering rate for an electron at
wavevector $p$ with spin $\sigma $ as:

\begin{eqnarray} \left. \frac {1}{\tau _{ee}} \right|_{p, \sigma } =  \frac
{2 \pi }{\hbar}  \sum_ {k, q} f_{k,\sigma' }\left( 1 - f_{k - q, \sigma' }
\right) \left( 1 - f_{p + q,\sigma }\right) {\left| \frac  {\langle k-q,
\sigma'; p+q, \sigma \left| V \right| k, \sigma';p, \sigma \rangle }
{\epsilon \left( q, \left( E_{p, \sigma } - E_{p+q, \sigma } \right) /\hbar
\right) } \right|} ^{2} \nonumber \\ \times \ \delta \left( E_{p+q,\sigma } +
E_{k-q, \sigma'} -  E_{p, \sigma } - E_{k, \sigma' } \right) \label{eq2}
\end{eqnarray}

where  $\langle k-q, \sigma';p+q, \sigma \left| V \right| k, \sigma' ;p,
\sigma \rangle = e^{2} F^{1D}_{ijkl}(q \times w)/(L
\epsilon_{0}\epsilon_{r})$ is the 1D  Coulomb interaction matrix element.
$F^{1D}_{ijkl}(q \times w)$ is the Coulomb Formfactor, calculated by
numerical integration, and $w$ is the wire width.  The dielectric function
$\epsilon \left( q, \left( E_{p, \sigma } - E_{p+q, \sigma } \right)/\hbar
\right)$ is calculated by integrating Ehrenreich's expression taking account
of finite temperature and the spin-split electron band structure including
non-parabolicity. The spin subband dependence  of the electron pair
scattering rates reported in this letter is caused by the Fermi population
factor $f_{k,\sigma' }\left( 1 - f_{k - q, \sigma' } \right) \left( 1 - f_{p
+ q,\sigma }\right) $.

Equation~\ref{eq2} is integrated numerically, taking a  quantum wire of width
$100\AA$ and square profile, with electron density $n=1.6 \times 10^{6}
cm^{-1}$ and at temperature $T=1.4K$, and with infinite confinement
potential. We include  a single `spin-up' band and a single `spin-down' band.
For the band dispersion  we use that of  bulk GaAs in the [110]  orientation
as a model. In an experimental quantum wire, the precise value of the
spin-splitting and the composition of the wave functions will depend on the
crystallographic details  and the microscopic electric fields present in the
structure.  Spin splitting is expected to be around $1 meV$ in GaAs near the
Fermi energy, thus limiting the predicted effects to temperatures below
around $10 K$. Higher temperatures are possible for materials with stronger
spin splitting.  Fig.~\ref{fig2}  shows, that the `forward'   pair scattering
rate (i.e. with partners near $+ {\bf k}_{F}$) increases with increasing
excess energy $\Delta $ for one `spin' subband (here `spin down') while it
remains suppressed for electrons in the opposite `spin' subband (here
`spin-up'). The `spin' subband dependence is almost absent for the weaker
`backward' scattering, i.e. with partners near $- {\bf k}_{F}$. Calculations
show that for hot electrons on balance a strong spin subband dependence of
the total scattering rates remains.

Fig.~\ref{fig3} demonstrates schematically the construction of such a quantum
wire spin polarizer. The wire length has to be less than the scattering
length for `spin-up' electrons (using the convention of the present Letter),
less than the probability for scattering with partners near  ${\bf k} \approx
- {\bf k}_{F}$ and longer than the scattering rate for electrons in the
`spin-down' subband. Calculation shows that this can be fulfilled in
$GaAs/Al_{x}Ga_{1-x}As$ based quantum wires, for excess energies $\Delta $ of
the order of $5 meV$, operating temperatures of $T = 4K$ or below, and wire
lengths in the $\mu m$-range. Acoustic phonon scattering is expected to be
weaker than  electron scattering effects up to at least $100K$, while we
expect that optical phonon scattering will destroy this effect above
approximately $100K$. Sufficiently high mobility is required, so that
impurity and roughness scattering are lower than electron-electron
scattering. Since in-built microscopic electric fields affect the
spin-splitting of the quantum wire, and since interface roughness can affect
microscopic electric field, it could also negatively affect the spin
polarization phenomena. Plasmon scattering is a possible loss mechanism
reducing efficiency and is neglected here.

So far we have assumed that the  background electron gas in the wire is
sufficiently coupled to the environment, so that its distribution is  not
disturbed by the injected electron beam.   The opposite limit is the case of
weak coupling of the background electrons in a section of the wire to the
surroundings, as demonstrated in Fig.~\ref{fig4}. In this case the injected
electron beam will flip background electrons between spin subbands with
unequal probability,  leading to unequal spin populations and a steady state
magnetization of the background electrons.

In summary, we have shown that an active electron spin polarizer can be
constructed from a quantum wire with spin splitting of electron bands. We
have calculated  the spin dependent differential electron pair scattering
rates as a function of electron excess energy. We have introduced a  further
related effect: a hot electron beam can induce spin polarization
(corresponding to a steady state magnetization) in a section of a quantum
wire, which is weakly coupled to the surroundings.  This work opens the
possibility of a large range of spin dependent experiments in microelectronic
structures.

\acknowledgments

The authors would like to express their gratitude to Professor Yasuhiko
Arakawa,  Dr. Yasushi Nagamune and Dr. Jun-Ichi Motohisa for helpful
discussions. Support of this work by the NTT Endowed Chair at RCAST and the
`Sakaki Quantum Wave Project'  of ERATO is gratefully acknowledged. GF would
like to express his gratitude to Dr. Bernard Jusserand for stimulating
discussions and for sending preprints prior to publication.

\narrowtext \figure{Diagram of intrasubband electron-electron pair scattering
processes in a quantum wire: the electron pair $\left( {\bf p} \uparrow ,
{\bf k} \downarrow \right)$ undergoes a pair scattering process, resulting in
the pair  $\left( {\bf p} \downarrow, {\bf k} \uparrow \right)$. In a
quantum wire only electron pairs with partners in  different `spin' subbands
can scatter. These  processes, introduced here, break   the electron phase
and have the character of    spin flip processes. \label{fig0}}

 \narrowtext \figure{This figure demonstrates that a stream of hot electrons
injected into a single mode quantum wire with an excess energy $\Delta $
above the Fermi level  may become spin polarized. {\bf (a)} A spin-up
electron injected with excess energy $\Delta $   and wave vector ${\bf
p}_{1}$ has  low scattering probability. {\bf (b)} An injected electron
injected into the `spin-down' subband with wave vector ${\bf k}_{2} - {\bf
q}_{2}$  has high scattering probability, represented by bold arrows.  After
such pair scattering, a hot electron with spin up at ${\bf p}_{1}$ is
produced, which has again low scattering probability. Thus for a stream of
hot electrons, the component  in the `spin-down' subband may be  converted
into  `spin-up' subband electrons, while the `spin-up' component  may pass
with little scattering.   \label{fig1}}

\narrowtext \figure{Differential electron-electron pair scattering rates as a
function of the wave vector of the scattering partner for the two spin
orientations, and for different excess energy $\Delta $  measured from the
Fermi surface.  For hot electrons ($\Delta > 0$),  electron pair scattering
rates may be orders of magnitude larger for one of the two spin subbands
(here spin-down) than for the opposite orientation. This effect is  due to
the Fermi population factors and the Pauli principle. (Divergences in the
calculated scattering rates at the point ${\bf q} =0$ have been eliminated
from the Figure. They are typical for 1D systems, are not expected to break
electron phase, and are not important for hot electrons). Note that the
scattering rates are only weakly spin subband dependent for `backward'
scattering, i.e. ${\bf q} \approx - 2 {\bf k}_{F}$. Calculations show that
for hot electrons a strong spin subband dependence of the total scattering
rates survives.      \label{fig2}}

\narrowtext \figure{Schematic design of an active quantum wire spin
polarizer. {\bf (a)} Outline of the conduction band potential of the wire
structure.  {\bf (b)} population of the lowest `spin-up' conduction subband,
{\bf (c)} population the lowest `spin-down' subband as a function of position
in the wire.     \label{fig3}}

\narrowtext \figure{Schematic diagram showing induced partial spin
polarization (corresponding to and induced steady state magnetization)
expected for a hot electron beam propagating through a section of a quantum
wire with spin split electron bands, in which the background electrons are
weakly coupled to the background. Shaded areas indicates steady state
distribution of passing hot electron beam, while the thick curves indicate
the population of the background electrons in the quantum wire --- note the
unequal number of populated `spin-up' and `spin-down' bands representing an
induced magnetization.   \label{fig4}}

 \end{document}